\documentclass[12pt]{article}

\usepackage[utf8]{inputenc}
\usepackage{geometry}
\usepackage{graphicx}
\usepackage{array} 
\usepackage{subfig}
\usepackage{slashed}
\usepackage{amsmath}
\usepackage{amsfonts}
\usepackage{amssymb}
\usepackage[cm]{fullpage}
\usepackage{cite}
\usepackage{epsfig}
\usepackage{comment}
\usepackage{hyperref}
\usepackage{tikz}
\usepackage{feynmp}
\usepackage{cancel}
\usepackage{mathrsfs}
\usepackage{mathtools}
\newcommand{\defeq}{\vcentcolon=}

\def\half{{1\over 2}}
\numberwithin{equation}{section}

\def\ip{${\mathscr I}^+$}

\def\e{{\epsilon}}
\def\cs{${\cal S}$}

 \def\p{\partial}
 \def\bz{{\bar z}}
 
\def\0{{(0)}}
\def\1{{(1)}}
\def\2{{(2)}}
 \def\cL{{\cal L}}

\def\ci{{\mathscr I}}

\def\<{\langle }
\def\>{\rangle }
\def\bw{{\bar w}}
\newcommand{\bea}{\begin{eqnarray}}
\newcommand{\eea}{\end{eqnarray}}
\newcommand{\be}{\begin{equation}}
\newcommand{\ee}{\end{equation}}
\newcommand{\ba}{\begin{align}}
\newcommand{\ea}{\end{align}}

\renewcommand{\epsilon}{\varepsilon}

   \makeatletter
  \let\over=\@@over \let\overwithdelims=\@@overwithdelims
  \let\atop=\@@atop \let\atopwithdelims=\@@atopwithdelims
  \let\above=\@@above \let\abovewithdelims=\@@abovewithdelims
\renewcommand\section{\@startsection {section}{1}{\z@}%
                                   {-3.5ex \@plus -1ex \@minus -.2ex}
                                   {2.3ex \@plus.2ex}%
                                   {\normalfont\large\bfseries}}

\renewcommand\subsection{\@startsection{subsection}{2}{\z@}%
                                     {-3.25ex\@plus -1ex \@minus -.2ex}%
                                     {1.5ex \@plus .2ex}%
                                     {\normalfont\bfseries}}

\newcommand{\bra}[1]{\left< #1 \right|}
\newcommand{\ket}[1]{\left| #1 \right>}

\newcommand{\beq}{\begin{equation}}
\newcommand{\eeq}{\end{equation}}
\newcommand{\beqa}{\begin{eqnarray}}
\newcommand{\eeqa}{\end{eqnarray}}
\newcommand{\beqar}{\begin{eqnarray*}}

\newcommand{\cl}{{\cal L}}

\newcommand{\cm}{{\cal M}}

\newcommand{\ve}{{\varepsilon}}

\def\[{\big[}
\def\]{\big]}

\newcommand{\bd}[1]{\begin{fmffile}{#1}\begin{fmfgraph*}}
\newcommand{\ed}{\end{fmfgraph*}\end{fmffile}}

\begin{document}
\begin{titlepage}
\unitlength = 1mm
\ \\
\vskip 3cm
\begin{center}

{ \LARGE {\textsc{ BMS Supertranslations and Weinberg's Soft Graviton Theorem}}}

\vspace{0.8cm}
Temple He, Vyacheslav Lysov, Prahar Mitra and Andrew Strominger

\vspace{1cm}

{\it  Center for the Fundamental Laws of Nature, Harvard University,\\
Cambridge, MA 02138, USA}

\begin{abstract}
Recently it was conjectured that a certain infinite-dimensional ``diagonal" subgroup of BMS supertranslations acting on 
past and future null infinity ($\ci^-$ and $\ci^+$) is an exact symmetry of the quantum gravity \cs-matrix, and an associated Ward identity was derived. In this paper we show that this supertranslation Ward identity is precisely equivalent to Weinberg's soft graviton theorem. Along the way we construct the canonical generators of  supertranslations at $\ci^\pm$, including the relevant soft graviton contributions. Boundary conditions at the past and future of $\ci^\pm$ and a correspondingly modified Dirac bracket are required.  The soft gravitons enter as  boundary modes and are manifestly the Goldstone bosons of spontaneously broken supertranslation invariance.

 \end{abstract}

\vspace{1.0cm}

\end{center}

\end{titlepage}

\pagestyle{empty}
\pagestyle{plain}

\def\vx{{\vec x}}
\def\p{\partial}
\def\po{$\cal P_O$}

\pagenumbering{arabic}

\tableofcontents

\section{Introduction}

Weinberg's soft graviton theorem \cite{steve} is a universal formula relating any \cs-matrix element in any quantum theory including gravity to a second \cs-matrix element which differs only by the addition of a graviton whose four-momentum is taken to zero. Remarkably, the formula is blind to the spin or any other quantum numbers of the asymptotic particles involved in the \cs-matrix element. 
   
It is often the case that universal formulae are explained by symmetries. Recently \cite{as}, it was conjectured that the quantum gravity \cs-matrix has an exact symmetry given by a certain infinite-dimensional ``diagonal" subgroup of the asymptotic supertranslation symmetries of Bondi, van der Burg, Metzner and Sachs (BMS) \cite{bms}. In this paper, we show that the universal soft graviton theorem of \cite{steve} is simply the Ward identity following from the diagonal BMS supertranslation symmetry of \cite{as}.
   
Put another way, it turns out that the deep discoveries made a half century ago about the structure of Minkowski scattering in theories with gravity by Weinberg and by BMS are equivalent, albeit phrased in very different languages. 

The Ward identities following from the diagonal BMS supertranslations were expressed in \cite{as} in terms of 
data at null infinity, namely the Bondi news representing gravitational radiation together with certain infrared modes. These are described in terms of their retarded times and positions on the asymptotic conformal sphere. The soft graviton theorem on the other hand is described \cite{steve} in terms of the scattering of momentum-space plane waves. The demonstration of this paper consists largely in transforming between these two different descriptions. 

In the course of our demonstration it is necessary to carefully define the physical phase spaces $\Gamma^\pm$ of gravitational modes at past and future null infinity ($\ci^-$ and $\ci^+$). $\Gamma^\pm$ must include, in addition to the Bondi news, all soft graviton degrees of freedom which do not decouple from the \cs-matrix. The latter we argue are constrained by boundary conditions at the boundaries of $\ci^\pm$. The soft modes can be viewed as living on these boundaries, and the  boundary conditions reduce their number by a crucial factor of 2.  The reduced space of modes may then be identified (from their transformation law)  as nothing but the Goldstone modes of spontaneously broken supertranslation invariance. The relevant physical phase spaces $\Gamma^\pm$ become simply the usual radiative modes plus the Goldstone modes.\footnote{This is the minimal phase space required for a good action of supertranslations. We have not ruled out the possibility of further soft modes and a larger phase space associated to local conformal symmetries\cite{bt,Banks:2003vp} which could  lie in components of the metric not considered here.} The boundary constraint entails a modification of the naive Dirac  bracket. After this modification canonical expressions for 
$T^\pm$ are given which generate supertranslations on all of $\Gamma^\pm$. While there has been much discussion of $T^\pm$ over the decades,  the construction of generators which act properly on the infrared as well as radiative modes is new.

This paper is organized as follows. In section 2 we present the full $\ci^\pm$ phase spaces $\Gamma^\pm$ (including the boundary condition), present the Dirac brackets and supertranslation generators $T^\pm$ and identify the soft gravitons as Goldstone modes. Section 3 reviews the proposed relation \cite{as} between \ip\ and $\ci^-$ near where they meet at spatial infinity, together with the diagonal supertranslations which preserve this relation and provide a symmetry of the \cs-matrix. Section 4 reviews the soft graviton theorem \cite{steve}. Section 5 describes the transformation between the asymptotic description of section 3 and the momentum space description of section 4. In section 6 we show that Weinberg's soft graviton theorem is the Ward identity following from diagonal supertranslation invariance.

We mainly consider only the case of pure gravity but expect the inclusion of massless matter or gauge fields to be straightforward. New elements may arise in theories which do not revert to the vacuum in the far past and future. We expect that parallel results apply to the gauge theory case \cite{as1}. Related results are in \cite{juan}. 

\section{ Supertranslation generators}
In this section we construct the physical phase space, the symplectic form (or equivalently the Dirac bracket)  and the canonical generators of supertranslations at $\ci^\pm$.

\subsection{Asymptotic vector fields}We consider asymptotically flat geometries in the finite neighborhood of Minkowski space defined in \cite{ck} and referred to in \cite{as} as CK spaces. These have a large-$r$ weak-field expansion near future null infinity (${\mathscr I}^+$) in retarded Bondi coordinates (see \cite{bt} for details) \begin{equation}
\begin{split}\label{metricexp}
ds^2 &= - du^2 - 2 du dr + 2 r^2 \gamma_{z {\bar z}} dz d{\bar z}  \\
&~~~~~~~~~~~~~~~~~~~~~~~~~~~~  + \frac{2m_B}{r} du^2 + r C_{zz} dz^2 + r C_{ {\bar z}{\bar z}} d{\bar z}^2 - 2 U_z du dz - 2 U_{\bar z} du d{\bar z}  + \cdots ,
\end{split}
\end{equation}
where 
\begin{equation}
\begin{split}\label{rel1}
U_z &= - \frac{1}{2} D^z C_{zz}.    \\
\end{split}
\end{equation}
The retarded time $u$ parameterizes the null generators of \ip\ and $(z,\bz)$ parameterize the conformal $S^2$. The Bondi mass aspect $m_B$ and $C_{zz}$ depend on $(u,z,\bz)$, $\gamma_{z {\bar z}}=\frac{2}{(1+z\bz)^2} $ is the round metric on  unit $S^2$ and $D_z$ is the $\gamma$-covariant derivative. Near past null infinity $\mathscr{I}^-$, CK spaces have a similar expansion in advanced Bondi coordinates 
\begin{equation}
\begin{split}
ds^2 &= - dv^2 + 2 dv dr + 2 r^2 \gamma_{z {\bar z}} dz d{\bar z}  \\
&~~~~~~~~~~~~~~~~~~~~~~~~~~~~ +  \frac{2m^-_B}{r} dv^2 + r D_{zz} dz^2 + r D_{ {\bar z}{\bar z}} d{\bar z}^2 - 2 V_z dv dz - 2 V_{\bar z} dv d{\bar z} + \cdots ,
\end{split}
\end{equation}
where
\begin{equation}
\begin{split}
V_z &=  \frac{1}{2} D^z D_{zz}. \\
\end{split}
\end{equation}
We denote the future (past) of \ip\ by  ${\mathscr{I}^+_+}$ (${\mathscr{I}^+_-}$), and the 
future (past) of $\ci^-$ by ${\mathscr{I}^-_+}$ (${\mathscr{I}^-_-}$). These comprise the boundary of $\ci$ $(\ci^+ + \ci^-)$. We also define the outgoing and incoming Bondi news by
\be N_{zz}\equiv\p_uC_{zz},~~~M_{zz}\equiv \p_vD_{zz}.\ee

BMS$^+$ transformations \cite{bms} are defined as the subgroup of diffeomorphisms which act nontrivially on the radiative data at \ip. These include the familiar Lorentz transformations and supertranslations. The latter are generated by the infinite family of vector fields\footnote{The subleading in $1 \over r$ terms depend on the coordinate condition: see \cite{bt}.}
\be \label{stn} f\p_u -{1\over r}(D^\bz f\p_\bz+D^z f \p_z ) 
+D^zD_z f\p_r,  \ee
for any function $f(z,\bz)$ on the $S^2$. 
BMS$^+$ acts on $C_{zz}$ according to  \be \label{dp} \cL_fC_{zz}=f\p_uC_{zz}-2D_z^2 f.\ee
 
Similarly BMS$^-$ transformations act on  $\ci^-$ and contain the supertranslations parameterized by $f^-(z,\bz)$
\be \label{ims} f^-\p_v +{1 \over r}(D^\bz f^-\p_\bz+D^z f^- \p_z )- D^zD_z f^-\p_r, \ee
under which
\be \label{dpw} \cL_{f^-}D_{zz}=f^-\p_vD_{zz}+2D_z^2 f^-.\ee

\subsection{Dirac brackets on $\ci$}
The Dirac bracket on the radiative modes (the non-zero modes of the Bondi news) at \ip\ was found in \cite{ash}
\be\label{db}  \{N_{\bz\bz}(u,z,\bz),N_{ww}(u',w,\bw)\}=-16\pi G \p_u\delta(u-u')\delta^2(z-w)\gamma_{z\bz},\ee
where $G$ is Newton's constant. 
The generator of BMS$^+$ supertranslations on these modes is \cite{ash,bt}
\begin{equation}
\begin{split}
\label{cok} T^+(f)&={1 \over 4\pi G}\int_{\ci^+_-}d^2z \gamma_{z\bz}fm_B \\
&= {1 \over 16\pi G}\int dud^2zf\big[ \gamma_{z\bz}N_{zz}N^{zz} +2 \p_u (\p_zU_\bz+\p_\bz U_z)\big], \\
\{T^+(f),N_{zz}\}&=f\p_uN_{zz},
\end{split}
\end{equation}
where in the second line we have used the constraints and assumed no matter fields. 

Of course
 BMS$^+$ transformations acting on the radiative modes alone do not comprise an asymptotic symmetry. One must act on a larger phase space $\Gamma^+$ including some non-radiative modes. The obvious guess is to identify this larger space with that parametrized by $C_{zz}$ itself, and define a bracket for all $(u,u')$ by integrating (\ref{db}) to  \be\label{zdb}  \{C_{\bz\bz}(u,z,\bz),C_{ww}(u',w,\bw)\}=8\pi G \Theta(u-u')\delta^2(z-w)\gamma_{z\bz},\ee
 where $\Theta(x)={\rm sign} (x)$. However if we use this we find, perhaps surprisingly, 
 \be\label{oo} \{T^+(f),C_{zz}\}=f\p_uC_{zz}-D_z^2f \neq \cL_fC_{zz}.\ee
 The inhomogeneous term is off by a factor of 2. So clearly either the bracket (\ref{zdb}) or the generator (\ref{cok}) is incorrect. This problem does not seem to have been addressed in the literature.

 Here we solve  this problem by motivating and imposing boundary conditions on $C_{zz}$  at the boundaries of \ip, and incorporating this boundary constraint into a modified Dirac bracket. Since the constraints apply only to the boundary degree of freedom, (\ref{zdb}) will be unaltered unless either $u$ or $u'$ is on the boundary. However this will turn out to give us exactly the missing factor of 2 in (\ref{oo})! The supertranslation invariant boundary conditions are
 \be \label{go}\[\p_zU_\bz-\p_\bz U_z\]_{\ci^+_\pm}= 0,\ee
 \be \label{sdc}N_{zz}|_{\ci^+_\pm}=0. \ee
 Equivalently the first condition may be written
 \be\label{sxz} \[D_z^2C_{\bz\bz}-D_\bz^2 C_{zz}\]_{\ci^+_\pm}=0. \ee
 This reduces the boundary degrees of freedom by a factor of two. 
It has a coordinate invariant expression in terms of the component of the Weyl tensor sometimes referred to as the magnetic mass aspect:
\be \text{Im}\; \Psi_2^\0|_{\ci^+_\pm}=0.\ee

There are two related motivations for this constraint besides the fact that it (as we will see momentarily) leads to a proper action of $T^+$. First, the boundary condition (\ref{go}) is obeyed by CK spaces \cite{ck}. Second, operator insertions of 
$\[\p_zU_\bz]^{\ci^+_+}_{\ci^+_-}$ and $\[\p_\bz U_z]^{\ci^+_+}_{\ci^+_-}$ correspond to soft gravitons and have non-vanishing \cs-matrix elements (due to Weinberg poles) even though they are pure gauge. Therefore they must be retained as part of the physical phase space. However these poles cancel in the difference $\[\p_zU_\bz]^{\ci^+_+}_{\ci^+_-}-\[\p_\bz U_z]^{\ci^+_+}_{\ci^+_-}$. Hence this combination  decouples from all \cs-matrix elements and should not be part of the physical phase space. Our constraint (\ref{go}) projects out these fully decoupled modes.

The general solution of the constraints (\ref{sxz}) can be expressed
\be C_{zz}|_{\ci^+_- }=  D^2_zC,\ee\be \label{xcv}  \int_{-\infty}^\infty du N_{zz}= D_z^2 N, \ee
where the boundary fields $C, N$ are real. We may then take as our coordinates on phase space the boundary and bulk fields\footnote{$C$ and $N$ each have four zero modes of angular momentum 0 and 1 which are projected out by $D_z^2$ and hence do not appear in the metric. They might be omitted from the definition of $\Gamma^+$ and do not play an important role in the present discussion. However we retain them for future reference:  as will become apparent below the  $C$ zero modes have an interesting interpretation as the spatial and temporal position of the geometry.}\be 
\Gamma^+\defeq \{ C(z,\bz), ~N(z,\bz),~C_{zz}(u,z,\bz),~C_{\bz \bz}(u,z,\bz)\}.\ee 
The arguments $u$ of the bulk fields  terms are restricted to non-boundary (i.e. finite) values only.  The bulk-bulk Dirac brackets remain (\ref{zdb}). A priori it is not obvious how one extends the bulk-bulk bracket (or equivalently the symplectic form) over all of $\Gamma^+$. We do so by first imposing (\ref{xcv}) as a relation between bulk-bulk and bulk-boundary brackets in the form
\be D_z^2\{N(z,\bz), C_{\bw\bw}(u,w,\bw)\}=\int_{-\infty}^\infty du' \{N_{zz}(u',z,\bz), C_{\bw\bw}(u,w,\bw)\},\ee
and then constraining the boundary-boundary bracket by continuity in the form
\be \label{ssq} D_\bw^2 \{N(z,\bz), C(w,\bw)\}= \lim_{u\to -\infty}\{N(z,\bz), C_{\bw\bw}(u,w,\bw)\}. \ee
The non-zero Dirac brackets following from the boundary constraints (\ref{sdc}), (\ref{sxz}) are then uniquely determined as\footnote{We note but do not pursue herein the interesting appearance of logarithms related to the four $C$ and $N$ zero modes. These are projected out by acting with $D_z^2$ and hence irrelevant to the supertranslation generators below. }
\begin{equation}
\begin{split}\label{dbc} 
 \{C_{\bz\bz}(u,z,\bz),C_{ww}(u',w,\bw)\}&=8\pi G \Theta(u-u')\delta^2(z-w)\gamma_{z\bz}, \\
 \{C(z,\bz),C_{ww}(u',w,\bw)\}&=  -  8G D_w^2(S \ln |z-w|^2) , \\
 \{N(z,\bz),C_{ww}(u',w,\bw)\}&=  16G D_w^2(S \ln |z-w|^2)  , \\
 \{N(z,\bz), C(w,\bw)\}&=  16G S \ln |z-w|^2 , 
  \end{split}
\end{equation}
where $u,u'$ are $not$ on the boundary and 
\begin{equation}
\begin{split}
 S\equiv {(z-w)(\bz-\bw) \over (1+z\bz)(1+w\bw)}.\;\;\; 
\end{split}
\end{equation}
$S$ is the sine-squared of the angle between $z$ and $w$ on the sphere and obeys
\bea
D_w^2(S \ln |z-w|^2)  &=&   \frac{S}{(z-w)^2}  ,\cr
 D_\bz^2D_{w}^2 (S\ln |z-w|^2) &=& \pi \gamma_{z\bz} \delta^2 (z-w) .\eea
 
 Similarly, on $\ci^-$, the constraints $\left[ \p_z V_\bz - \p_\bz V_z \right]_{\ci^-_\pm} = 0$ can be solved by\be D_{zz}|_{\ci^-_+}=  D^2_z D, \;\;\;\;  \int_{-\infty}^\infty dv M_{zz} = D_z^2 M.\ee
 The coordinates on the phase space at $\ci^-$ can then be taken as
 \be 
\Gamma^- \defeq \{ D(z,\bz), ~M(z,\bz),~D_{zz}(v,z,\bz),~D_{\bz \bz}(v,z,\bz)\},\ee 
where $v$ is \textit{not} on the boundary. The non-zero Dirac brackets are 
\begin{equation}
\begin{split}\label{dbd}
 \{D_{\bz\bz}(v,z,\bz),D_{ww}(v',w,\bw)\}&=8\pi G \Theta(v-v')\delta^2(z-w)\gamma_{z\bz}, \\
 \{D(z,\bz),D_{ww}(v',w,\bw)\}& = 8G  D_w^2(S \ln |z-w|^2), \\
  \{M(z,\bz),D_{ww}(v',w,\bw)\}& =16 G  D_w^2(S \ln |z-w|^2), \\
 \{M(z,\bz) ,D (w,\bw)\}&= 	 16GS \ln |z-w|^2,  
 \end{split}
\end{equation}
where $v,v'$ are \textit{not} on the boundary.

The demand of continuity (\ref{ssq}) is not as innocuous as it looks because we see from (\ref{dbc}), (\ref{dbd}) that other brackets (in particular $\{N_{zz},C_{ww}\}$)  are $not$ continuous as $u$ is taken to the boundary.  We have not ruled out the possibility that there are inequivalent extensions of the symplectic form on the radiative phase space to all of $\Gamma^\pm$ corresponding to inequivalent quantizations of the boundary sector. In an action formalism, this could arise from different choices of boundary terms.  However an {\it a posteriori} justification of our choice is, as we now show, that it leads to a realization of supertranslations as a canonical transformation on $\Gamma^\pm$. 
 \subsection{Canonical generators}
The  supertranslation generator may now be written in terms of bulk and boundary fields as 
\bea \label{dok} T^+(f)={1 \over 4\pi G}\int_{\ci^+_-}d^2z \gamma_{z\bz}fm_B={1 \over 16\pi G}\int dud^2zf\gamma_{z\bz}N_{zz}N^{zz} - {1 \over 8\pi G}\int d^2z\gamma^{z\bz}fD_z^2D_\bz^2N ,\eea
where  the integral over  infinite $u$ in the first term is the Cauchy principal value. 
Using the brackets   (\ref{dbc})  one finds 
\begin{equation}\label{ssx}
\begin{split}
\{T^+(f),N_{zz}\}&=f\p_uN_{zz}, \\
\{T^+(f),C_{zz}\}&= f \p_uC_{zz}-2D_z^2f, \\
\{T^+(f),N\}&=0, \\
\{T^+(f),C\}&=-2f,
\end{split}
\end{equation}
as desired. 

Similarly on $\ci^-$,
\begin{equation}
\label{dxk}
T^-(f^-) ={1 \over 16\pi G}\int dvd^2zf^-\gamma_{z\bz}M_{zz}M^{zz}  + {1 \over 8\pi G}\int d^2z\gamma^{z\bz}f^-D_z^2D_\bz^2M,  
\end{equation}
and
\begin{equation}
\begin{split}
\{T^-(f^-),M_{zz}\}&=f^-\p_vM_{zz}, \\
\{T^-(f^-),D_{zz}\}&= f^-\p_vD_{zz}+2D_z^2f^- , \\
\{T^-(f^-),M\}&= 0, \\
\{T^-(f^-),D\}&= 2f^- , 
\end{split}
\end{equation}
as desired. 

At the quantum level supertranslations do not leave the usual in or out vacua invariant.  Acting with $T^+$, the last term in  (\ref{dok}) is linear in the graviton field operator and creates a new state with a soft graviton. The new state has energy degenerate with the out vacuum but different angular momentum. Hence supertranslation symmetry is spontaneously broken in the usual vacuum. The last line of (\ref{ssx}) clearly identifies $-\half C$ as the Goldstone mode associated with this symmetry breaking. It is conjugate to the soft graviton zero mode 
$N$.

In conclusion the construction of a generator of supertranslations on $\ci^\pm$ is possible but subtle and requires a careful analysis of the zero mode structure and boundary conditions on the boundaries of $\ci^\pm$.

\section{Supertranslation invariance of the \cs-matrix}

In this section we summarize the supertranslation invariance of the \cs-matrix conjectured in \cite{as} as well as the associated Ward identity.

The first step is to understand how \ip\ and $\ci^-$ may be linked near spatial infinity. In the conformal compactification of asymptotically flat spaces, the sphere at spatial infinity is the boundary of a point $i^0$. Null generators of $\ci$ in the conformal compactification of asymptotically flat spaces run from $\ci^-$ to \ip\ through $i^0$. We label all points  lying on the same such generator with the same value of $(z, \bz)$. This gives an `antipodal' identification of points on the conformal spheres at $\ci^-$ with those on \ip.  
For CK spaces one may identify geometric 
data on ${\mathscr{I}^+_-}$ with that at ${\mathscr{I}^-_+}$ via the continuity condition \cite{as}
\be \label{cnvt} C_{zz}|_{\mathscr{I}^+_-}=-D_{zz}|_{\mathscr{I}^-_+},  \ee
or equivalently
\be C(z,\bz)=-D(z,\bz).\ee

In \cite{as} it was conjectured that the ``diagonal" subgroup of BMS$^+\times$BMS$^-$ which preserves the continuity condition 
(\ref{cnvt}) is an exact symmetry of both classical gravitational scattering and the quantum gravity \cs-matrix. The diagonal supertranslation generators are those which are constant on the null generators of $\ci$, $i.e.$  
\be f^-(z,\bz)=f(z,\bz).\ee
The conjecture states that \cs-matrix obeys
\be \label{fs}T^+(f){\cal S}-{\cal S}T^-(f)=0.\ee 
A Ward identity is then derived by taking the matrix elements of (\ref{fs}) between states with $n$ incoming ($m$ outgoing) particles at 
$z_k^{\text{in}}$ ($z_k^{\text{out}}$) on the conformal sphere at $\ci$.  These carry energies $E_k^{\text{in}}$ ($E_k^{\text{out}}$), where
\begin{equation}
\begin{split}
\sum_{k=1}^m E_k^{\text{out}} = \sum_{k=1}^n E_k^{\text{in}} 
\end{split}
\end{equation}
by total energy conservation. We denote the out and in  states by  $\bra{z_1^{\text{out}}, \cdots}$ and  $ \ket{z_1^{\text{in}},  \cdots}$. Choosing $f(w,\bw)={1 \over z-w}$,  it was shown that the matrix element of (\ref{fs}) between such states implies
\begin{equation}\label{sstw}
\begin{split}
\bra{z_1^{\text{out}}, \cdots} \colon P_z {\cal S} \colon \ket{z_1^{\text{in}},  \cdots} = \bra{z_1^{\text{out}},   \cdots}    {\cal S}   \ket{z_1^{\text{in}},  \cdots}  \left[ \sum_{k=1}^m \frac{E_k^{\text{out}}}{z - z_k^{\text{out}}}  - \sum_{k=1}^n \frac{E_k^{\text{in}}}{z - z_k^{\text{in}}}  \right],
\end{split}
\end{equation}
where the $\colon~\colon$ denotes time-ordering and the ``soft graviton current"  is defined by 
\begin{equation}
\begin{split}\label{pzcurr}
P_z \equiv \frac{1}{2G} \left( \int^\infty_{-\infty}dv\p_vV_z - \int^\infty_{-\infty} du\p_uU_z \right) .
\end{split}
\end{equation}
 Since $P_z$ involves zero-frequency integrals over $\ci^\pm$ it creates and annihilates soft gravitons with a certain $z$-dependent wave function. The supertranslation Ward identity (\ref{sstw}) relates \cs-matrix elements with and without insertions of the soft graviton current. It can also easily be seen \cite{as} that (\ref{sstw}) implies the general Ward identities following from (\ref{fs}) for an   arbitrary  function $f(z,\bz)$.

\section{The soft graviton theorem}
In this section, we specify our conventions and briefly review Weinberg's derivation of  the soft graviton theorem for the simplest case of a free massless scalar. For more details and general spin see \cite{steve}. 
 
Einstein gravity coupled to a free massless scalar is described by the action
\begin{equation}
\begin{split}
S = - \int d^4 x \sqrt{-g} \left[ \frac{2 }{ \kappa^2}R + \frac{1}{2} g^{\mu\nu} \p_\mu \phi \p_\nu \phi \right] ,
\end{split}
\end{equation}
where $\kappa^2 =  32 \pi G $. In the weak field perturbation expansion  $g_{\mu\nu} = \eta_{\mu\nu} + \kappa h_{\mu\nu}$ and the relevant leading terms are
\begin{equation}
\begin{split}\label{gravpreact}
\cl_{\text{grav}}&=  - \frac{2}{\kappa^2} R = - \frac{1}{2} \p_\sigma h_{\mu\nu} \p^\sigma h^{\mu\nu} + \frac{1}{2} \p_\mu h \p^\mu h  +  \p^\mu   h_{\mu\nu}\p_\rho h^{\nu\rho} -  \p_\mu h^{\mu\nu}  \p_\nu   h + \cdots , \\
\cl_s &= -\frac{1}{2} \sqrt{-g} g^{\mu\nu} \p_\mu \phi \p_\nu \phi = - \frac{1}{2} \p^\mu \phi \p_\mu \phi +  \frac{1}{2} \kappa h^{\mu\nu} \left[    \p_\mu \phi \p_\nu \phi - \frac{1}{2}   \eta_{\mu\nu}   \p^\sigma \phi \p_\sigma \phi \right]  + \cdots .
\end{split}
\end{equation}
In harmonic gauge $\p^\mu h_{\mu\nu} = \frac{1}{2} \p_\nu h$ the Feynman rules take the form (see \cite{css})
\begin{equation}
\begin{split}
\bd{photonprop}(50,30)
\fmfset{arrow_len}{2.5mm}
\fmfleft{i}
\fmfright{o}
\fmf{dbl_curly,label=$p$,label.side=left}{i,o}
\fmfv{label=$\mu\nu$}{i}
\fmfv{label=$\alpha\beta$}{o}
\ed 
&~~~~~~~~~ \raisebox{13 pt}{\begin{normalsize}$=  \frac{1}{2} \left( \eta_{\mu\alpha} \eta_{\nu\beta} + \eta_{\mu\beta} \eta_{\nu\alpha} - \eta_{\mu\nu} \eta_{\alpha\beta} \right) \frac{ - i}{p^2 - i \e} ,  $\end{normalsize}} \\
\bd{scalarprop}(50,30)
\fmfset{arrow_len}{2.5mm}
\fmfset{arrow_len}{2.5mm}
\fmfleft{i}
\fmfright{o}
\fmf{fermion,label=$p$,label.side=left}{i,o}
\ed 
&~~~~~~~~~ \raisebox{13 pt}{\begin{normalsize}$= \frac{ - i}{ p^2 - i \e} , $\end{normalsize}} \\ & \\
\bd{3point}(50,40)
\fmfset{arrow_len}{2.5mm}
\fmfleft{i}
\fmfright{o}
\fmftop{t}
\fmf{fermion,label=$p_1$,label.side=right}{i,g}
\fmf{fermion,label=$p_2$,label.side=right}{g,o}
\fmf{dbl_curly,label=$p$,label.side=left}{t,g}
\fmfv{label=$\mu\nu$}{t}
\ed 
&~~~~~~~~~ \raisebox{30 pt}{\begin{normalsize}$= \frac{i \kappa}{2} \left( p_{1\mu} p_{2\nu} + p_{1\nu} p_{2\mu} - \eta_{\mu\nu} p_1 \cdot p_2 \right) . $\end{normalsize}}
\end{split}
\end{equation}
Now, consider an on-shell amplitude involving $n$ incoming (with momenta $p_1, \cdots, p_n$) and $m$ outgoing (with momenta $p'_1, \cdots,p'_m$) massless scalars represented by the diagram
\\
\begin{equation}
\begin{split}
\bd{amp}(80,80)
\fmfleft{i1,i2,i3}
\fmfright{o1,o2,o3}
\fmf{fermion}{i1,g}
\fmf{fermion}{i2,g}
\fmf{fermion}{i3,g}
\fmf{fermion}{g,o1}
\fmf{fermion}{g,o2}
\fmf{fermion}{g,o3}
\fmfv{decor.shape=circle,decor.filled=shaded,decor.size=20}{g}
\fmfv{label=$p'_1$}{o1}
\fmfv{label=$\cdots$}{o2}
\fmfv{label=$p'_m$}{o3}
\fmfv{label=$\cdots$}{i2}
\fmfv{label=$p_1$}{i1}
\fmfv{label=$p_n$}{i3}
\ed
\end{split}
\end{equation}
\\
Consider the same amplitude with an additional outgoing soft graviton of momentum $q$ and polarization $\ve_{\mu\nu}(q)$ satisfying the gauge condition $q^\mu \ve_{\mu\nu} = \frac{1}{2} q_\nu \ve^\mu{}_\mu$. The dominant diagrams in the soft $q\to 0$ limit are \\
\begin{equation}
\begin{split}
\bd{ampwithgrav}(80,80)
\fmfleft{i1,i2,i3}
\fmfright{o1,o2,o3,o5}
\fmf{fermion}{i1,g}
\fmf{fermion}{i2,g}
\fmf{fermion}{i3,g}
\fmf{fermion}{g,o1}
\fmf{fermion}{g,o2}
\fmf{fermion}{g,o3}
\fmf{dbl_curly}{o5,g}
\fmfv{decor.shape=circle,decor.filled=shaded,decor.size=20}{g}
\fmfv{label=$p'_1$}{o1}
\fmfv{label=$\vdots$}{o2}
\fmfv{label=$p'_m$}{o3}
\fmfv{label=$q$}{o5}
\fmfv{label=$\vdots$}{i2}
\fmfv{label=$p_1$}{i1}
\fmfv{label=$p_n$}{i3}
\ed ~~~~~~ \raisebox{37pt}{$=~~ \sum\limits_{k=1}^m $}~~~
\bd{ampwithgrav1}(100,80)
\fmfleft{i1,i2,i3}
\fmfright{o1,o2,o3,o4}
\fmf{fermion}{i1,g}
\fmf{fermion}{i2,g}
\fmf{fermion}{i3,g}
\fmf{fermion}{g,o1}
\fmf{fermion}{g1,o2}
\fmf{fermion}{g,g1}
\fmf{dbl_curly}{o3,g1}
\fmf{fermion}{g,o4}
\fmfv{decor.shape=circle,decor.filled=shaded,decor.size=20}{g}
\fmfv{label=$p'_1$}{o1}
\fmfv{label=$p'_k$}{o2}
\fmfv{label=$q$}{o3}
\fmfv{label=$p'_m$}{o4}
\fmfv{label=$\vdots$}{i2}
\fmfv{label=$p_1$}{i1}
\fmfv{label=$p_n$}{i3}
\ed  ~~\raisebox{37pt}{$~~ + ~~~ \sum\limits_{k=1}^n  $}~~~~~
 \bd{ampwithgrav2}(100,80)
\fmfleft{o1,o2,o4}
\fmfright{i1,i2,i3,o3}
\fmf{fermion}{g,i1}
\fmf{fermion}{g,i2}
\fmf{fermion}{g,i3}
\fmf{fermion}{o1,g}
\fmf{fermion}{o2,g1}
\fmf{fermion}{g1,g}
\fmf{dbl_curly}{g1,o3}
\fmf{fermion}{g,o4}
\fmfv{decor.shape=circle,decor.filled=shaded,decor.size=20}{g}
\fmfv{label=$p_1$}{o1}
\fmfv{label=$p_k$}{o2}
\fmfv{label=$q$}{o3}
\fmfv{label=$p_n$}{o4}
\fmfv{label=$\vdots$}{i2}
\fmfv{label=$p'_1$}{i1}
\fmfv{label=$p'_n$}{i3}
\ed
\end{split}
\end{equation}\\
Additional diagrams with the external graviton attached to internal lines cannot develop soft poles\cite{steve}. 
The contribution of these diagrams to the near-soft amplitude is 
\begin{equation}\label{zzk}
\begin{split}
\cm_{\mu\nu} (q, p'_1, \cdots,p'_m,p_1,\cdots,p_n) &= \sum\limits_{k=1}^m \cm  (p'_1, \cdots,p'_k+q, \cdots, p'_m,p_1,\cdots,p_n) \frac{- i}{ \left( p'_k + q \right)^2 - i \e} \\
&~~~~~~~~  \times \left[ \frac{i \kappa}{2} \left( p'_{k\mu} ( p'_k+q)_\nu + p'_{k\nu} ( p'_k+q)_\mu  - \eta_{\mu\nu} p'_k \cdot \left( p'_k + q\right) \right) \right]  \\
&~~~~ +  \sum\limits_{k=1}^n \cm  (p'_1, \cdots, p'_m,p_1,\cdots,p_k+q,\cdots,p_n) \frac{ - i}{ \left( p_k - q \right)^2 - i \e} \\
&~~~~~~~~ \times \left[  \frac{i \kappa}{2} \left( p_{k\mu} ( p_k - q)_\nu + p_{k\nu} ( p_k-q)_\mu  - \eta_{\mu\nu} p_k \cdot \left( p_k - q \right) \right) \right] .
\end{split}
\end{equation}
The soft graviton theorem is the leading term in $q$-expansion:
\begin{equation}
\begin{split} \label{sgt}
 \cm_{\mu\nu} (q, p'_1, \cdots,p'_m,p_1,\cdots,p_n) = \frac{\kappa}{2} \left[  \sum\limits_{k=1}^m  \frac{  p'_{k\mu}  p' _{k\nu}}{p'_k \cdot q  }  -   \sum\limits_{k=1}^n  \frac{   p_{k\mu}  p_{k\nu} }{  p_k \cdot q}  \right]  \cm  (p'_1, \cdots, p'_m,p_1,\cdots,p_n) ,
\end{split}
\end{equation}
where $q \to 0$. While we reviewed the derivation here for a massless scalar, note that the prefactor in square brackets is a universal soft factor and does not depend on the spin of the matter particles. Moreover the expression is actually gauge invariant. Under a gauge transformation 
$\delta \ve^{\mu \nu}=q^\mu\Lambda^\nu +q^\nu\Lambda^\mu$ one finds 
\be \delta \ve^{\mu \nu} \cm_{\mu\nu}= \kappa \Lambda^\mu \left[ \sum\limits_{k=1}^m   p'_{k\mu} - \sum\limits_{k=1}^n    p_{k\mu}    \right]  \cm=0 \ee
by momentum conservation. Hence (\ref{sgt}) is valid in any gauge. 

\section{From momentum to asymptotic position space}
The supertranslation Ward identity (\ref{sstw}) is expressed in terms of field operator $P_z$ integrated along fixed-angle null generators of $\ci$.
Weinberg's soft graviton theorem (\ref{zzk}) is expressed  in terms of momentum eigenmodes of the field operators. In this section, in order to compare the two, we transform the field operator between these two bases. 

Momentum eigenmodes in Minkowski space are usually described in the flat coordinates \be\label{fm}ds^2=-dt^2+d\vec x \cdot d\vec x.\ee
These flat coordinates are related to the retarded coordinates (\ref{metricexp}) by the transformation
\bea 
t&=&u+r,\cr
x^1+ix^2&=&{2rz\over 1+z\bz},\cr
x^3&=&{r(1-z\bz)\over 1+z\bz},\eea
with $\vec x=(x^1,x^2,x^3)$ obeying  $\vec x \cdot \vec x =r^2$. 
At late times and large $r$  the wave packet for a massless particle with spatial momentum centered around $\vec p$
becomes localized on the conformal  sphere near  the point 
\be\label{mpr} \vec{p} = \omega {\hat x} \equiv \omega {\vec x \over r}={\omega \over 1+z\bz}(z+\bz,-iz+i\bz,1-z\bz), \ee
where $\vec p\cdot\vec p=\omega^2$. Hence, the momentum of massless particles may be equivalently characterized by $(\omega, z,\bz)$
or $p^\mu$. 

At late times $t\to\infty$ the gravitational field becomes free and can be approximated by the mode expansion 
\begin{equation}
\begin{split}\label{modexp}
h^{\text{out}}_{\mu\nu}(x) = \sum\limits_{\alpha=\pm} \int \frac{ d^3q}{(2\pi)^3} \frac{1}{2 \omega_q} \left[ \ve^{\alpha*}_{\mu\nu} ({ \vec q})a^{\text{out}}_\alpha ({\vec q}) e^{i q \cdot x} + \ve^\alpha_{\mu\nu}({\vec q}) a^{\text{out}}_\alpha ({\vec q})^\dagger   e^{- i q \cdot x} \right],
\end{split}
\end{equation}
where $q^0 = \omega_q = | {\vec q}|$, $\alpha=\pm$ are the two helicities and
\be\label{rrd}
\[a^{\text{out}}_\alpha ({\vec q}), a^{\text{out}}_\beta ({\vec{q'}})^\dagger \]= \delta_{\alpha\beta} (2\omega_q)   (2\pi)^3  \delta^3 \left( {\vec q} - {\vec {q'}} \right). 
\ee
The outgoing gravitons with momentum $q$ and polarization $\alpha$ as in the amplitude (\ref{zzk}) correspond to final-state insertions of $a^{\text{out}}_\alpha ({\vec q})$.

Let us now take $(\omega_q,w,\bw)$ to parameterize the graviton four-momentum
\begin{equation}
\begin{split}\label{gravmom}
q^\mu = \frac{\omega_q}{1 + w {\bar w}} \left( 1 + w {\bar w} , w + {\bar w} , - i \left( w - {\bar w}\right), 1 - w {\bar w}  \right).
\end{split}
\end{equation}
The polarization tensors may be written $\ve^{\pm\mu\nu}=\ve^{\pm\mu}\ve^{\pm\nu}$ with
\begin{equation}
\begin{split}\label{pols1}
{\bf \ve}^{+\mu}( {\vec q} ) &= \frac{1}{\sqrt{2}} \left( {\bar w}, 1, - i, - {\bar w} \right),  \\
{\bf \ve}^{-\mu}({\vec q} ) &= \frac{1}{\sqrt{2}} \left( w , 1,   i, - w  \right). \\
\end{split}
\end{equation}
These obey  $\ve^{\pm\mu\nu}q_\nu= \ve^{\pm\mu}{}_\mu=0$  and 
\be \label{vv}\e_z^+ \left({\vec q} \right) = \p_z x^\mu\ve^+_\mu \left( \vec{q} \right) = \frac{ \sqrt{2} r {\bar z} \left( {\bar w} - {\bar z} \right) }{ \left( 1 + z {\bar z} \right)^2 } ,~~~~\e_z^- \left({\vec q} \right) = \p_z x^\mu\ve^-_\mu \left( \vec{q} \right)  = \frac{ \sqrt{2} r \left( 1 + w {\bar z} \right)}{ \left( 1 + z {\bar z} \right)^2 } .\ee
In retarded Bondi coordinates, it follows from \eqref{metricexp} that on ${\mathscr I}^+$ \begin{equation}
\begin{split}
C_{zz}(u,z,\bz) = \kappa \lim_{r\to\infty} \frac{1}{r} h^{\text{out}}_{zz}(r,u,z,\bz).
\end{split}
\end{equation}
Using $h_{zz} = \p_z x^\mu \p_z x^\nu h_{\mu\nu}$ and the mode expansion \eqref{modexp}, we find
\begin{equation}
\begin{split}
C_{zz} &= \kappa \lim_{r\to\infty} \frac{1}{r} \p_z x^\mu \p_z x^\nu   \sum\limits_{\alpha=\pm} \int \frac{ d^3q}{(2\pi)^3} \frac{1}{2 \omega_q} \left[ \ve^{\alpha*}_{\mu\nu} ({ \vec q})a^{\text{out}}_\alpha ({\vec q}) e^{- i \omega_q u - i \omega_q r \left( 1 - \cos\theta \right) } + h.c. \right] .
\end{split}
\end{equation}
where $\theta$ is the angle between $\vec{x}$ and $\vec{q}$. The integrand oscillates rapidly at large $r$ with  stationary points at $\theta=0,\pi$. The stationary phase approximation to the integral over the momentum-space sphere then gives  
\begin{equation}
\begin{split}\label{Czzmodexp}
C_{zz} &=     - \frac{ i \kappa  }{ 4 \pi^2 \left(1 + z {\bar z} \right)^2 }  \int_0^\infty d \omega_q   \left[  a^{\text{out}}_+ (\omega_q {\hat x }) e^{- i \omega_q u } -    a^{\text{out}}_- (\omega_q {\hat x })^\dagger e^{i \omega_q u } \right], 
\end{split}
\end{equation}
where the contribution from the $\theta=\pi$ stationary point vanishes in the large $r$ limit. Defining
\be \begin{split}\label{eq1sgop}
N_{zz}^\omega(z,{\bar z}) &\equiv \int_{-\infty}^\infty du e^{ i \omega u} \p_u C_{zz}, \\
\end{split}
\end{equation}
and using \eqref{Czzmodexp}, we find
\begin{equation}
\begin{split}\label{nzzeq}
N_{zz}^\omega(z,{\bar z}) &=  - \frac{ \kappa  }{2 \pi \left(1 + z {\bar z} \right)^2 } \int_0^\infty d \omega_q \omega_q  \left[  a^{\text{out}}_+ (\omega_q {\hat x })  \delta(\omega_q - \omega ) +  a^{\text{out}}_- (\omega_q {\hat x })^\dagger  \delta(\omega_q + \omega) \right].
\end{split}
\end{equation}
When $\omega >0$ ($\omega<0$), only the first (second) term contributes and we find
\begin{equation}
\begin{split}
N^\omega_{zz}(z,{\bar z}) &=  - \frac{ \kappa \omega a^{\text{out}}_+ (\omega {\hat x })  }{2 \pi \left(1 + z {\bar z} \right)^2 }, \\
N_{zz}^{-\omega}(z,{\bar z})  &=  - \frac{ \kappa \omega a^{\text{out}}_- (\omega {\hat x })^\dagger  }{2 \pi \left(1 + z {\bar z} \right)^2 },
\end{split}
\end{equation}
where we have taken $\omega > 0$. In the case of the zero mode, we will define it in a hermitian way
\be N_{zz}^0\equiv \lim_{\omega \to 0^+} \half( N_{zz}^{\omega}+ N_{zz}^{-\omega}).\ee It follows that \begin{equation}
\begin{split}
N_{zz}^0(z,{\bar z}) &=  - \frac{ \kappa  }{4 \pi \left(1 + z {\bar z} \right)^2 } \lim_{\omega \to 0^+}  \left[  \omega  a^{\text{out}}_+ (\omega {\hat x })  + \omega  a^{\text{out}}_- (\omega  {\hat x })^\dagger   \right] .
\end{split}
\end{equation}

A parallel construction is possible on  ${\mathscr I}^-$. Defining
\begin{equation}
\begin{split}\label{eq2sgop}
M_{zz}^\omega(z,{\bar z}) \equiv \int_{-\infty}^\infty dv e^{  i \omega v} \p_v D_{zz} ,\\
\end{split}
\end{equation}
we find for $\omega > 0$
\begin{equation}
\begin{split}
M_{zz}^\omega(z,{\bar z}) & =  -  \frac{ \kappa \omega   a^{\text{in}}_+ (  \omega  {\hat x })  }{2 \pi  \left( 1 + z {\bar z} \right)^2} ,\\
M_{zz}^{-\omega}(z,{\bar z})  & =   - \frac{ \kappa \omega   a^{\text{in}}_- (  \omega  {\hat x})^\dagger  }{2 \pi  \left( 1 + z {\bar z} \right)^2}, 
\end{split}
\end{equation}
where $a^{\text{in}}_\pm$ and $a^{\text{in}\dagger}_\pm$ annihilate and create incoming gravitons on ${\mathscr I}^-$. At $\omega = 0$, 
\begin{equation}\label{bum}
\begin{split}
M_{zz}^0(z,{\bar z}) &\equiv - \frac{\kappa}{4\pi \left( 1+z {\bar z}\right)^2  }  \lim\limits_{\omega \to 0^+} \left[  \omega   a^{\text{in}}_+ ( \omega {\hat x})  + \omega a^{\text{in}}_-(\omega {\hat x})^\dagger   \right] .
\end{split}
\end{equation}

From \eqref{eq1sgop} and \eqref{eq2sgop}  we have also 
\begin{equation}
\begin{split}
N_{zz}^0(z,{\bar z}) &= D_z^2 N , \\
M_{zz}^0(z,{\bar z}) &=    D_z^2 M.  
\end{split}
\end{equation}
Defining 
\begin{equation}
\begin{split}\label{eq37}
{\cal O}_{zz} &\equiv N_{zz}^0(z,{\bar z})  + M_{zz}^0(z,{\bar z}) =   D_z^2 N + D_z^2 M ,
\end{split}
\end{equation}
the soft graviton current \eqref{pzcurr} can be written
\begin{equation}
\begin{split}\label{pzeq}
P_z = \frac{1}{2 G} \left( V_z|^{{\mathscr I}^-_+}_{{\mathscr I}^-_-} - U_z |^{{\mathscr I}^+_+}_{{\mathscr I}^+_-} \right) = \frac{1}{4 G} \gamma^{z {\bar z}} \p_{\bar z}{\cal O}_{zz}.
\end{split}
\end{equation}

\section{Soft graviton theorem as a Ward identity}
Equations (\ref{bum})-(\ref{pzeq}) express the soft graviton current $P_z$ in terms of standard momentum space creation and annihilation operators. Amplitudes involving the latter are given by Weinberg's soft graviton theorem. In this section we simply plug this in and reproduce the supertranslation Ward identities. 

We denote an \cs-matrix element of $m$ incoming and $n$ outgoing particles by
\begin{equation}
\begin{split}
\bra{z_1^{\text{out}}, \cdots}   {\cal S}   \ket{ z_1^{\text{in}}, \cdots} ,
\end{split}
\end{equation}
where the in (out) momenta are parametrized by $z^{\text{in}}$ ($z^{\text{out}}$) as in  (\ref{mpr}).
We now consider the \cs-matrix element $\bra{z_1^{\text{out}}, \cdots} \colon {\cal O}_{zz} {\cal S} \colon \ket{ z_1^{\text{in}}, \cdots}$ with a time ordered insertion. Using (\ref{bum}) and \eqref{eq37}, this can be written as
\begin{equation}
\begin{split}
\bra{z_1^{\text{out}}, \cdots} \colon {\cal O}_{zz} {\cal S} \colon \ket{ z_1^{\text{in}}, \cdots} &= - \frac{\kappa}{4\pi \left( 1 + z {\bar z} \right)^2 }\lim_{\omega \to 0^+} \left[ \omega  \bra{z_1^{\text{out}}, \cdots}     a^{\text{out}}_+ (\omega {\hat x})       {\cal S}   \ket{ z_1^{\text{in}}, \cdots} \right. \\
&\left. ~~~~~~~~~~~~~~~~~~~~~~~~~~~~~~~~ + \omega  \bra{z_1^{\text{out}}, \cdots}   {\cal S}     a^{\text{in}}_- (  \omega {\hat x})^\dagger \ket{ z_1^{\text{in}}, \cdots}   \right] .
\end{split}
\end{equation}
Here, we have used the fact that $a^{\text{out}}_-(\omega {\hat x} )^\dagger$ ($a^{\text{in}}_+(\omega {\hat x}))$ annihilates the out (in) state for $\omega \to 0$.\footnote{This holds even if for example the initial  state contains soft gravitons because of the factor of $\omega$ in (\ref{rrd}).} The first term is the \cs-matrix element with a single outgoing positive helicity soft graviton with spatial momentum $\omega {\hat x}$, while the second term is the \cs-matrix element with a single incoming negative helicity soft graviton also with spatial momentum $ \omega {\hat x}$. The two amplitudes are equal, and we get
\begin{equation}
\begin{split}\label{eq111}
\bra{z_1^{\text{out}}, \cdots} \colon {\cal O}_{zz} {\cal S} \colon \ket{ z_1^{\text{in}}, \cdots} &= - \frac{\kappa}{2\pi \left( 1 + z {\bar z} \right)^2 }\lim_{\omega \to 0^+} \left[ \omega  \bra{z_1^{\text{out}}, \cdots}     a^{\text{out}}_+ (\omega {\hat x})       {\cal S}   \ket{ z_1^{\text{in}}, \cdots}  \right].
\end{split}
\end{equation}
The soft graviton theorem \eqref{sgt} with a positive helicity outgoing graviton reads
\begin{equation}
\begin{split}
& \lim_{\omega \to 0} \left[ \omega  \bra{z_1^{\text{out}}, \cdots}     a^{\text{out}}_+ ({\vec q})       {\cal S}   \ket{ z_1^{\text{in}}, \cdots}  \right] \\
&~~~~~~~~~~~~~~~~~~~~~ = \frac{\kappa}{2}\lim_{\omega \to 0^+}  \left[  \sum\limits_{k=1}^m  \frac{ \omega \left[ p'_k \cdot \ve^+(q)\right]^2  }{p'_k \cdot q  }  -   \sum\limits_{k=1}^n  \frac{\omega  \left[ p_k \cdot \ve^+(q) \right]^2  }{  p_k \cdot q }  \right] \bra{z_1^{\text{out}}, \cdots}     {\cal S}   \ket{ z_1^{\text{in}}, \cdots}  .
\end{split}
\end{equation}
Using the parametrization of the momenta discussed earlier
\begin{equation}
\begin{split}
p_k^\mu &=  E^{\text{in}}_k \left(1 ,  \frac{ z^{\text{in}}_k + {\bar z}^{\text{in}}_k}{ 1 + z^{\text{in}}_k {\bar z}^{\text{in}}_k }, \frac{- i \left(  z^{\text{in}}_k - {\bar z}^{\text{in}}_k \right) }{ 1 + z^{\text{in}}_k {\bar z}^{\text{in}}_k }, \frac{1- z^{\text{in}}_k {\bar z}^{\text{in}}_k}{1+ z^{\text{in}}_k {\bar z}^{\text{in}}_k} \right) ,\\
p'^\mu_k &=  E^{\text{out}}_k \left(1 ,  \frac{ z^{\text{out}}_k + {\bar z}^{\text{out}}_k}{ 1 + z^{\text{out}}_k {\bar z}^{\text{out}}_k }, \frac{- i \left(  z^{\text{out}}_k - {\bar z}^{\text{out}}_k \right) }{ 1 + z^{\text{out}}_k {\bar z}^{\text{out}}_k }, \frac{1- z^{\text{out}}_k {\bar z}^{\text{out}}_k}{1+ z^{\text{out}}_k {\bar z}^{\text{out}}_k} \right)  ,\\
q^\mu &=  \omega \left(1 ,  \frac{ z + {\bar z}}{ 1 + z {\bar z} }, \frac{- i \left(  z - {\bar z} \right) }{ 1 + z {\bar z} }, \frac{1- z {\bar z}}{1+ z {\bar z}} \right) ,\\
{\bf \ve}^{+\mu}(q) &= \frac{1}{\sqrt{2}} \left( {\bar z}, 1, - i, - {\bar z} \right),  
\end{split}
\end{equation}
and \eqref{eq111}, we find
\begin{equation}
\begin{split}
\bra{z_1^{\text{out}}, \cdots} \colon {\cal O}_{zz} {\cal S} \colon \ket{ z_1^{\text{in}}, \cdots} &=  \frac{8  G }{  \left( 1 + z {\bar z} \right) }  \bra{z_1^{\text{out}}, \cdots}     {\cal S}   \ket{ z_1^{\text{in}}, \cdots}  \\
& \times \left[  \sum\limits_{k=1}^m  \frac{ E_k^{\text{out}} \left( {\bar z} - {\bar z}^{\text{out}}_k \right)}{ \left( z - z^{\text{out}}_k \right) \left( 1 + z^{\text{out}}_k {\bar z}^{\text{out}}_k \right) } -    \sum\limits_{k=1}^n  \frac{ E_k^{\text{in}} \left( {\bar z} - {\bar z}^{\text{in}}_k \right)}{ \left( z - z^{\text{in}}_k \right) \left( 1 + z^{\text{in}}_k {\bar z}^{\text{in}}_k \right) }  \right]  .
\end{split}
\end{equation}
Now, using \eqref{pzeq}, we can relate the insertion of $P_z$ to that of ${\cal O}_{zz}$. 
\begin{equation}
\begin{split}
\bra{z_1^{\text{out}}, \cdots} \colon P_z {\cal S} \colon \ket{ z_1^{\text{in}}, \cdots} &= \frac{1}{4G} \gamma^{z {\bar z}} \p_{\bar z} \bra{z_1^{\text{out}}, \cdots} \colon {\cal O}_{zz} {\cal S} \colon \ket{ z_1^{\text{in}}, \cdots}  \\
&=    \bra{z_1^{\text{out}}, \cdots}     {\cal S}   \ket{ z_1^{\text{in}}, \cdots}   \left[  \sum\limits_{k=1}^m  \frac{ E_k^{\text{out}}  }{   z - z^{\text{out}}_k  } -    \sum\limits_{k=1}^n   \frac{ E_k^{\text{in}}  }{   z - z^{\text{in}}_k  } \right]   \\
&~~~~~~  + \bra{z_1^{\text{out}}, \cdots}     {\cal S}   \ket{ z_1^{\text{in}}, \cdots}   \left[  \sum\limits_{k=1}^m  \frac{ E_k^{\text{out}}  {\bar z}^{\text{out}}_k }{ 1 + z^{\text{out}}_k {\bar z}^{\text{out}}_k   }  -    \sum\limits_{k=1}^n  \frac{ E_k^{\text{in}}  {\bar z}^{\text{in}}_k }{ 1 + z^{\text{in}}_k {\bar z}^{\text{in}}_k   }  \right]  .
\end{split}
\end{equation}
The very last square bracket vanishes due to total momentum conservation.
We then have
\begin{equation}
\begin{split}
 \bra{z_1^{\text{out}}, \cdots} \colon P_z {\cal S} \colon \ket{ z_1^{\text{in}}, \cdots}  =     \bra{z_1^{\text{out}}, \cdots}     {\cal S}   \ket{ z_1^{\text{in}}, \cdots}   \left[  \sum\limits_{k=1}^m  \frac{ E_k^{\text{out}}  }{   z - z^{\text{out}}_k  } -    \sum\limits_{k=1}^n   \frac{ E_k^{\text{in}}  }{   z - z^{\text{in}}_k  } \right]   ,
\end{split}
\end{equation}
which reproduces exactly the supertranslation Ward identity \eqref{sstw} derived in \cite{as}. We can also run the above argument backwards to show that this supertranslation Ward identity implies Weinberg's soft graviton theorem.

\appendix

\section*{Acknowledgements}
We are grateful to G. Barnich and G. Compere for useful conversations and to J. Maldacena and  A. Zhiboedov for discussions and for sharing their unpublished work \cite{juan} containing related results. This work was supported in part by NSF grant 1205550  and the Fundamental Laws Initiative at Harvard.

\end{document}